\documentclass[aps,prl,twocolumn,groupedaddress,superscriptaddress,showpacs]{revtex4-1}

\usepackage{amsmath, amsthm, amssymb}

\usepackage{graphicx}
\usepackage{psfrag}
\usepackage{dsfont}
\usepackage{tabularx}

\usepackage{xcolor}
\definecolor{mydarkgreen}{rgb}{0.0,0.5,0.0}

\usepackage[linktocpage=true,
  colorlinks=true, 
  pdfborder={0 0 0},
  linkcolor=blue,
  citecolor=red,
  filecolor=yellow,
  urlcolor=mydarkgreen,
  bookmarks,
  pdftitle={FT-RDMFT: exchange only},
  pdfauthor={Tim Baldsiefen},
 pdfkeywords={Reduced density matrix functional theory,electron gas},
]{hyperref} 

\newcommand{\tb}[1]{\textbf{#1}}
\newcommand{\tr}[1]{\textup{tr}\{#1\}}
\newcommand{\eq}[1]{(\ref{#1})}

\newcommand{\ket}[1]{{|#1\rangle}}
\newcommand{\bra}[1]{{\langle#1|}}

\newcommand{\nn}{\nonumber}
\newcommand{\veps}{\varepsilon}

\newcommand{\abref}[1]{#1}

\newcolumntype{x}[1]{>{\centering\arraybackslash\hspace{0pt}}p{#1}}

\newlength{\figurewidth}

\begin{document}


\title{Reduced density matrix functional theory at finite temperature. II. \\Application to the electron gas: Exchange only}


\author{Tim Baldsiefen}
\affiliation{Institut f\"ur Theoretische Physik, Freie Universit\"at Berlin, Arnimallee 14, D-14195 Berlin, Germany}
\affiliation{Max-Planck-Institut f\"ur Mikrostrukturphysik, Weinberg 2, D-06112 Halle, Germany}

\author{F. G. Eich}
\affiliation{Institut f\"ur Theoretische Physik, Freie Universit\"at Berlin, Arnimallee 14, D-14195 Berlin, Germany}
\affiliation{Max-Planck-Institut f\"ur Mikrostrukturphysik, Weinberg 2, D-06112 Halle, Germany}

\author{E. K. U. Gross}
\affiliation{Max-Planck-Institut f\"ur Mikrostrukturphysik, Weinberg 2, D-06112 Halle, Germany}


\date{\today}

\begin{abstract}
Using the newly introduced theory of finite-temperature reduced density matrix functional theory \cite{Baldsiefen_al_1.2012}, we apply the first-order approximation to the homogeneous electron gas. We consider both collinear spin states as well as symmetry broken states describing planar spin spirals and investigate the magnetic phase diagram as well as the temperature-dependence of the single particle spectra.
\end{abstract}


\maketitle


\section{Introduction}
An accurate theoretical description of the homogeneous electron gas (HEG) poses a central, though not yet completely solved, problem in quantum mechanics. Highly accurate Monte-Carlo calculations of the groundstate energy of the HEG with collinear spin-magnetism form the cornerstone of the highly successful local-spin-density approximations LSDA \cite{Ceperley_Alder.1980,Perdew_Wang.1992} within density functional theory (DFT). At finite temperature, the equilibrium thermodynamic variables are in principle accessible via path-integral quantum Monte-Carlo calculations \cite{Ceperley.1995}. However, the fermionic sign problem renders an accurate calculation of the equilibrium properties of the HEG at low temperatures almost impossible \cite{Troyer_Wiese.2005}. Accordingly, a wide variety of different approaches were developed. These include the introduction of approximate model interactions \cite{Zong_Lin_Ceperley.2002,Conduit_Green_Simons.2009}, the utilization of the dielectric formulation (employing various approximations, including the hypernetted chain approximation \cite{Dharma-Wardana_Perrot.1982,Iyetomi_Ichimaru.1986}, the modified convolution approximation \cite{Ichimaru_Tanaka.1987, Tanaka_Ichimaru.1989} and the equation-of-motion approach of Singwi, Tosi, Land and Sj\"olander \cite{Singwi_Tosi_Land_Sjolander.1968,Tanaka_Ichimaru.1986,Schweng_Boehm.1993}), the mapping of quantum systems to classical systems at finite temperature \cite{Perrot_Dharma-Wardana.2000,Dharma-Wardana_Perrot.2003}, and the utilization of finite-temperature many body perturbation theory (FT-MBPT), including non-diagrammatic local field corrections to the random phase approximation \cite{Perrot.1979,Gupta_Rajagopal_2.1980,Dharma-Wardana_Taylor.1981,Gupta_Rajagopal.1982,Perrot_Dharma-Wardana.1984,Kanhere_Panat_Rajagopal_Callaway.1986,Dandrea_Ashcroft_Carlsson.1986}.

As an alternative to these approaches, in this work we will consider the theoretical framework of finite-temperature reduced density matrix functional theory (FT-RDMFT) \cite{Baldsiefen_al_1.2012}. This allows a variational treatment of the grand potential of an arbitrary quantum system, employing the 1-reduced density matrix (1RDM). Compared to the conceptually similar framework of FT-DFT,  this finite-temperature version of RDMFT is capable of describing the interacting kinetic energy as well as the exchange energy excactly and only the correlation contributions to the interaction energy and to the entropy need to be approximated. FT-RDMFT also allows the calculation of the grand potential of a HEG subject to a nonlocal external potential. 

The general purpose of this work is twofold. Firstly, we introduce the general concept of FT-RDMFT on the basis of the first-order functional, neglecting correlation. This paves the way for the treatment of correlation in Part III of this work \cite{Baldsiefen_al_3.2012}, which in the framework of FT-RDMFT consists of the inclusion of an additional functional. 

Secondly, as this first-order functional yields the finite-temperature Hartree-Fock (FT-HF) solution, we can investigate the temperature dependence of the different grand potential contributions, the magnetic phase diagram, and the single-particle spectra for the HEG in FT-HF for both collinear as well as spin spiral configurations, therefore extending previous FT-HF results \cite{Hong_Mahan.1994,Hong_Mahan.1995}.

\setlength{\figurewidth}{0.9\linewidth}

\section{Theoretical Framework}
\subsection{FT-RDMFT}
We will now shortly review the most important concepts from our article on the foundations of FT-RDMFT \cite{Baldsiefen_al_1.2012}.

In the present work, we treat grand canonical quantum ensembles, i.e. systems in contact with a particle and a heat bath. For such systems the main thermodynamic variable is the grand potential $\Omega=E-\mu N-1/\beta S$. $E$ describes the internal energy, $N$ the particle number, $S$ the entropy of the system, and $\beta=1/(k_BT)$ where $k_B$ is Boltzmann's constant. The chemical potential $\mu$ and the temperature $T$ govern the coupling to the particle and heat baths respectively. A general state of a quantum mechanical system is described by a statistical density operator (SDO) $\hat D$.
\begin{align}
  \hat D&=\sum_iw_i\ket{\Psi_i}\bra{\Psi_i}\quad,w_i\geq0,\sum_iw_i=1\label{eq.sdo},
\end{align}
where $\{\ket{\Psi_i}\}$ forms a basis of the Fock space under consideration. The grand potential and entropy expressed as functionals of the SDO are
\begin{align}
  \Omega[\hat D]&=\tr{\hat D(\hat H-\mu\hat N-1/\beta\ln\hat D}\\
  S[\hat D]&=-\tr{\hat D\ln\hat D},
\end{align}
where $\hat H$ is the Hamiltonian of the system and $\hat N$ is the particle number operator. The SDO minimizing $\Omega[\hat D]$ is given by $\hat D_{eq}=e^{-\beta(\hat H-\mu\hat N)}/\tr{e^{-\beta(\hat H-\mu\hat N)}}$. On the basis of this variational principle, Mermin \cite{Mermin.1965} showed that there is a one-to-one correspondence between an external local one-particle potential and the corresponding equilibrium density $n_{eq}$ rendering a finite-temperature version of DFT possible. However, the kinetic energy as well as big parts of the interaction, including the exchange energy, are not known as explicit functionals of the density and have to be approximated. It is known, on the other hand, that the kinetic energy and the exchange energy can be treated exactly by using the 1RDM of the system. The resulting theory, RDMFT, showed some success at zero temperature e.g. in describing the fundamental gap in molecules as well as solids, including transition metal oxides \cite{Lathiothakis_al.2009,Sharma_al.2008,Baldsiefen.2010}. The density $n(\tb r)$ and the 1RDM $\gamma(\tb r,\tb r')$ of a quantum system in a state described by $\hat D$ are given by
\begin{align}
  \gamma_{\sigma\sigma'}(\tb r,\tb r')&=\tr{\hat D\hat\psi_{\sigma'}^+(\tb r')\hat\psi_\sigma(\tb r)}\label{eq.1rdm}\\
  n_\sigma(\tb r)&=\gamma_{\sigma\sigma}(\tb r,\tb r)\label{eq.n},
\end{align}
with $\{\hat\psi_\sigma(\tb r)\}$ being the common field operators. By construction, the 1RDM is hermitean and can be diagonalized. We have shown in Part I \cite{Baldsiefen_al_1.2012} that for some special cases, the 1RDM can be seperated into two distinct spin contributions. As the symmetries we are considering in this work allow such a seperation, we will write the 1RDM in spectral representation as
\begin{align}
  \gamma_{\sigma\sigma'}(\tb r,\tb r')&=\delta_{\sigma\sigma'}\sum_in_{i\sigma}\phi_{i\sigma'}^*(\tb r')\phi_{i\sigma}(\tb r),
\end{align}
where, due to L\"owdin \cite{Loewdin.1955}, we call the eigenvalues $\{n_{i\sigma}\}$ and eigenstates $\{\phi_{i\sigma}(\tb x)\}$ occupation numbers (ON) and natural orbitals (NO) respectively. It was shown by Coleman \cite{Coleman.1963} that if the NOs form a basis and the ONs fulfill $0\leq n_{i\sigma}\leq1,\sum_{i\sigma} n_{i\sigma}=N$ then the 1RDM is ensemble-$N$-representable, i.e. it corresponds to a statistical density operator of the form \eq{eq.sdo}.

An immediate consequence of Mermin's proof together with Eqs. \eq{eq.1rdm} and \eq{eq.n} is the one-to-one correspondence between $\hat D_{eq}$ and $\gamma_{eq}$ for the case of local external potentials. In the case of nonlocal external potentials one needs to use Gilbert's \cite{Gilbert.1975} theorem to show that the one-to-one correspondence between $\hat D_{eq}$ and $\gamma_{eq}$ still prevails. Hence one can describe the equilibrium properties of a grand canonical ensemble by means of functionals of the 1RDM. Furthermore, in contrast to zero-temperature RDMFT there now exists a noninteracting Kohn-Sham (KS)-system following from the simple invertibility of the Fermi-Dirac-distribution \cite{Dirac.1926}.
\begin{align}
  n_{i\sigma}&=\frac{1}{1+e^{\beta(\varepsilon_{i\sigma}-\mu)}}\\
  \varepsilon_{i\sigma}-\mu&=\frac1\beta\ln\left(\frac{1-n_{i\sigma}}{n_{i\sigma}}\right)\label{eq.fdinv}
\end{align} 

The existence of a Kohn-Sham system allows the construction of a perturbative expansion of the grand potential functional. The first-order functional for the grand potential is then given by

\begin{multline}
  \Omega[\gamma]=\Omega_k[\gamma]+V_{ext}[\gamma]-\mu N[\gamma]-1/\beta S_0[\gamma]\\
  +\Omega_H[\gamma]+\Omega_x[\gamma],\label{eq.gp1.def}
\end{multline}
where the general forms of the different functional contributions are given in Part I \cite{Baldsiefen_al_1.2012}.

Although the perturbative method is explicitly derived on the premise of describing a grand canonical ensemble, we have shown \cite{Baldsiefen_al_1.2012} that under the assumption of being in the thermodynamic limit, the functionals for the thermodynamic variables of a grand canonical ensemble coincide with the ones describing a canonical one. In the following, we are therefore able to use our method of FT-RDMFT for the description of an electron gas in a canonical ensemble. The equilibrium state is then found by a minimization of the free energy functional $F[\gamma]$ as defined as
\begin{align}
  F[\gamma]&=\Omega[\gamma]+\mu N[\gamma].
\end{align}

%
%
%

\section{Homogeneous Electron Gas}
A central model system for the theoretical description of many particle quantum systems is the HEG, an extensive review of which can be found in Ref. \cite{Giuliani_Vignale}. The density $\rho$ of the HEG is defined by the Wigner-Seitz radius $r_s$, i.e. the radius of a sphere of constant density which contains one electron
\begin{align}
  r_s&=\left(\frac{3}{4\pi\rho}\right)^{1/3}.
\end{align}
The characteristic energy of a system at density $r_s$ is the Fermi energy $\veps_F$ with the corresponding Fermi temperature $T_F$
\begin{align}
  \veps_F&=\frac{3^{\frac43}\pi^\frac23}{2^{\frac73}r_s^2}\\
  T_F&=\frac{\veps_F}{k_B}\approx 5.83\cdot 10^6 r_s^{-2}.
\end{align}

By choosing a basis set of certain symmetry, one restricts the domain of minimization of the free energy functional to states respecting this symmetry of the system. In the following we will focus on two different symmetries, the first one describing collinear spins and the second one describing a chiral spin symmetry.

\subsection{Collinear spins}
The NOs describing a system with collinear spin symmetry are plane waves. Assuming furthermore that the charge distribution is uniform, i.e. that we are not in the Wigner-crystal phase, the 1RDM becomes
\begin{align}
  \gamma_{\sigma\sigma'}(\tb r- \tb r')&=\delta_{\sigma\sigma'}\int\frac{d^3k}{(2\pi)^3}n_\sigma(\tb k)e^{i\tb k\cdot(\tb{r}-\tb r')}.
\end{align}
The polarization is then defined as
\begin{align}
  \xi&=\frac{N^\uparrow-N^\downarrow}{N^\uparrow+N^\downarrow}.
\end{align}
with $N^{\sigma}=(2\pi)^{-3}\int d^3kn_\sigma(\tb k)$. The functionals for the kinetic, exchange, and entropic contributions to the free energy are then given as
\begin{align}
  \Omega_k[\gamma]&=\frac1\rho\sum_\sigma\int\frac{d^3k}{(2\pi)^3}n_\sigma(\tb k)\frac{k^2}{2}\label{eq.heg.coll.k}\\
  \Omega_x[\gamma]&=-\frac{1}{2\rho}\sum_{\sigma}\int\frac{dk_1^3}{(2\pi)^3}\int\frac{dk_2^3}{(2\pi)^3}\nn\\
  &\hspace{30mm}n_\sigma(\tb k)n_\sigma(\tb k')\frac{4\pi}{(\tb k-\tb k')^2}\label{eq.heg.coll.x}\\
  S_0[\gamma]&=-\frac{1}{\rho}\sum_{\sigma}\int\frac{d^3k}{(2\pi)^3}\Big(n_\sigma(\tb k)\ln(n_\sigma(\tb k))\nn\\
  &\hspace{2cm}+(1-n_\sigma(\tb k))\ln(1-n_\sigma(\tb k))\Big)\label{eq.heg.coll.s}
\end{align}

In the numerical treatment, we assume the ONs to be constant in small volumes $V_i$ around the k-points $\tb k_i$.
Eqs. \eq{eq.heg.coll.k} - \eq{eq.heg.coll.s} then transform into sums and we arrive at the final expression for the free energy functional:
\begin{align}
  F[\{n_{i\sigma}\}]&=\Omega_k[\{n_{i\sigma}\}]+\Omega_x[\{n_{i\sigma}\}]-1/\beta S_0[\{n_{i\sigma}\}]\label{eq.heg.f}.
\end{align}
The kinetic energy $\Omega_k$, the exchange energy $\Omega_x$, and the noninteracting entropy $S_0$ are given by
\begin{align}
  \Omega_k[\{n_{i\sigma}\}]&=\sum_{i,\sigma} n_{i\sigma} t_i\\
  \Omega_x[\{n_{i\sigma}\}]&=-\frac12\sum_{i,j,\sigma}n_{i\sigma}n_{j\sigma}K_{i,j}\label{eq.heg.ex}\\
  S_0[\{n_{i\sigma}\}]&=-\sum_{i,\sigma}\Big(n_{i\sigma}\ln(n_{i\sigma})\nn\\
  &\hspace{2cm}+(1-n_{i\sigma})\ln(1-n_{i\sigma})\Big)\omega_i\label{eq.heg.s0},
\end{align}
where
\begin{align}
  t_i&=\frac{1}{\rho}\int_{V_i}\frac{d^3k}{(2\pi)^3}\frac{k^2}{2}\label{eq.weight.k}\\
  K_{i,j}&=\frac{1}{\rho}\int_{V_i}\frac{dk_1^3}{(2\pi)^3}\int_{V_j}\frac{dk_2^3}{(2\pi)^3}\frac{4\pi}{(\tb k-\tb k')^2}\label{eq.weight.x}\\
  \omega_i&=\frac{1}{\rho}\int_{V_i}\frac{d^3k}{(2\pi)^3}\label{eq.weight.w}.
\end{align}
We are now able to investigate the magnetic phase transitions of a HEG at finite temperature.

\begin{figure}[t!]
  \includegraphics[width=\figurewidth]{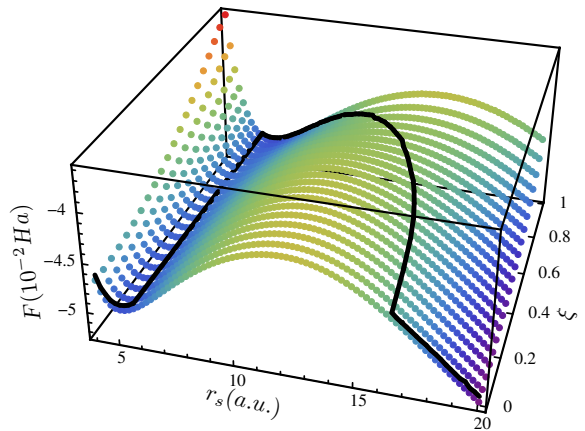}
  \caption{Free energy w.r.t. polarization $\xi$ and Wigner-Seitz radius $r_s$ at $T=5000K$. The black line denotes the equilibrium free energies for a fixed $r_s$. At $r_s\approx 5.8a.u.$ an instantaneous phase transition between paramagnetic and ferromagnetic phases takes place. An increase in $r_s$ then leads to a continuous transition back to the paramagnetic state.}
  \label{fig.heg.surf}
\end{figure}

As an example, we plot the free energy at $T=5000K$ versus polarization and density of the HEG in Figure \ref{fig.heg.surf}. The positions of the free energy minima for each $r_s$ are denoted by the black line. When reducing the density, i.e. increasing $r_s$, we encounter an instantaneous phase transition between a paramagnetic (vanishing polarization) and ferromagnetic (complete polarization) equilibrium phase.

As is well known, at zero temperature, the transition from a paramagnetic to a ferromagnetic state can be explained on the basis of the $r_s^{-2}$ and $r_s^{-1}$ dependencies of kinetic and exchange energy respectively. At small $r_s$, the kinetic energy, which favours a paramagnetic configuration, is dominant. Increasing $r_s$ then increases the effect of the exchange energy, which favours the ferromagnetic state. At some critical density $r_c$ the effect of the exchange contribution finally overcomes the kinetic one and a magnetic quantum phase transition occurs. That this transition is instantaneous cannot be deduced from this simple argument and one would have to examine the explicit form of both kinetic as well as exchange energy. To explain the finite-temperature behaviour, one can essentially repeat the previous considerations, however, with the entropy contribution included. The entropy generally favours a completely disordered state which in our situation means a paramagnetic one. Also, the entropy has no explicit density dependence. Therefore, for increasing $r_s$ the entropy, compared to kinetic and exchange contributions, becomes more and more dominant. This explains why for increasing $r_s$ the paramagnetic state will again become favourable at finite temperature. These simple arguments explain the existence of the phase transitions in Figure \ref{fig.heg.surf}. Our results show that the first-order functional of FT-RDMFT yields an instantaneous transition between paramagnetic and ferromagnetic configurations which is not the case in real systems where the quantum phase transition is of second order. This result has to be attributed to the first-order approximation as Monte-Carlo results for the HEG show a continuous change of the order parameter, i.e. the polarization, at zero temperature \cite{Perdew_Wang.1992,Zong_Lin_Ceperley.2002}.

\begin{figure}[t!]
  \psfrag{ty}{$\xi_{min}$}
  \psfrag{tx}{$r_s(a.u.)$}
  \psfrag{t0}[bl][bl][1][0]{$T=0K$}
  \psfrag{t5}[bl][bl][1][0]{$T=5000K$}
  \psfrag{t8}[bl][bl][1][0]{$T=8000K$}
  \psfrag{t10}[bl][bl][1][0]{$T=10000K$}
  \includegraphics[width=\figurewidth]{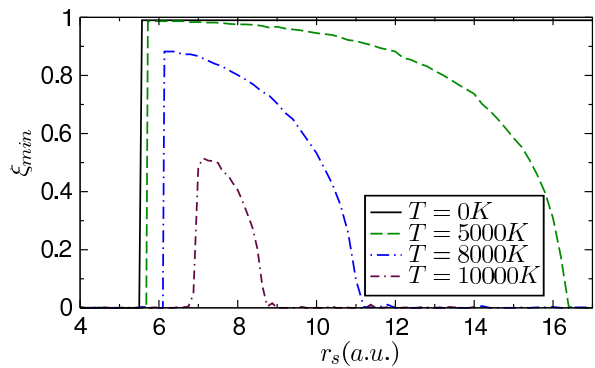}
  \caption{Equilibrium polarization $\xi_{min}$ of the HEG as function of $r_s$ for different temperatures $T$. Above $T=6000K$ there is no fully polarized equilibrium state.}
  \label{fig.heg.minpol}
\end{figure}

Since the entropic term in the free energy has the temperature as a prefactor, one expects the ferromagnetic phase to vanish faster with increasing temperature. Figure \ref{fig.heg.minpol} shows the validity of this argument. Interestingly, the nature of the phase transitions does not change, even at high temperatures.

\begin{figure}[t!]
  \centering
  \includegraphics[width=\figurewidth]{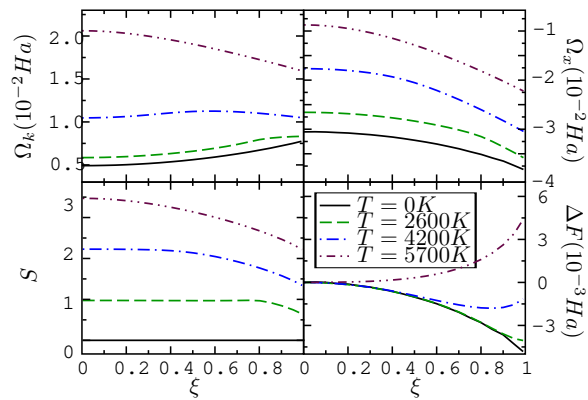}
  \caption{Clockwise from bottom left: entropy, kinetic energy, exchange energy, and free energy at $r_s=15 a.u.\ (T_F=2590K)$ versus the polarization $\xi$ of the HEG for different temperatures $T$. The reduced temperatures are $t={0.0,1.0,1.6,2.2}$. For the free energy, we plot $\Delta F(\xi)=F(\xi)-F(\xi=0)$.}
  \label{fig.heg.e-rs}
\end{figure}

In order to study the various contributions to the free energy seperately we have plotted kinetic energy, exchange contribution and entropy in Figure \ref{fig.heg.e-rs}.
We see that the entropy and exchange contributions always show a monotonically decreasing behaviour w.r.t. an increase in the polarization $\xi$. The kinetic energy however, which is known at zero temperature to be monotonically increasing with $\xi$, actually becomes decreasing for high values of $T/T_F$.

\begin{figure}[t!]
  \centering
  \includegraphics[width=\figurewidth]{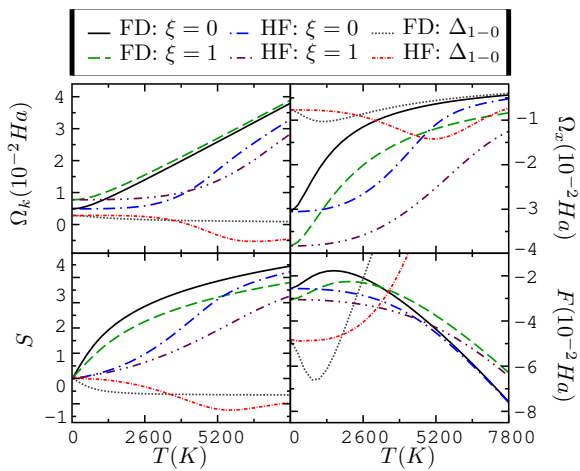}
  \caption{Clockwise from bottom left: entropy, kinetic energy, exchange energy, and free energy for the \abref{HF} and the noninteracting functionals at $r_s=15 a.u.\ (T_F=2590K)$ and $\xi=(0,1)$ versus the temperature $T$ of the \abref{HEG}. The black dotted and the grey shaded lines denote the differences between the ferro- and paramagnetic configurations. The free energy differences are scaled by 10. The behaviour of the \abref{HF}-curves can be explained by different ``{}effective''{} Fermi temperatures.}
  \label{fig.applications.mbpt-fd-hf}
\end{figure}

As we will see in the following, this somewhat counterintuitive effect is due to the fact that the exchange contribution hinders the temperature-induced smoothening of the momentum distributions and stronger so for the ferromagnetic configuration. To elucidate this argument we choose a density which will yield a ferromagnetic solution at zero temperature ($r_s=15a.u.$). Thermodynamic variables for both paramagnetic and ferromagnetic configurations as functions of the temperature are shown in Figure \ref{fig.applications.mbpt-fd-hf}. The curves denoted by ``FD'' correspond to the \abref{FT-MBPT} expressions, i.e. the \abref{FT-RDMFT} functionals applied to Fermi-Dirac momentum distributions with the appropriate temperature. The \abref{FT-RDMFT} functional is then minimized to give the curves denoted by ``HF''. The differences of the energies for ferromagnetic and paramagnetic configurations are then included as the ``$\Delta_{1-0}$'' curves.
\begin{align}
  \Delta_{1-0}&=\Omega(\xi=1)-\Omega(\xi=0)
\end{align}
We see that in the case of a noninteracting system, i.e. for the ``FD'' curves, the kinetic energies of paramagnetic and ferromagnetic configurations approach each other but do not cross. The fact that they converge for high temperatures can qualitatively be understood by the concept of different ``effective'' Fermi temperatures $T_F^*$ of the configurations. For the paramagnetic configuration at zero temperature the ONs occupy two Fermi spheres of radii $k^u_{F\uparrow}=k^u_{F\downarrow}=k_F$, one for each spin channel. Because in the ferromagnetic situation the ONs are restriced to only one spin channel there will also be only one Fermi sphere with increased radius $k^p_{F\uparrow}=2^\frac13k_F\ (k^p_{F\downarrow}=0)$. Because the kinetic energies are proportional to $\sum_\sigma (k_{F\sigma})^2$ this explains the favourisation of the paramagnetic configuration at zero temperature. An increase in temperature will now lead to a smoothening of the Fermi sphere, i.e. the momentum distributions, and therefore to an overall increase of the kinetic energy. The quickness of the smoothening is determined mainly by the characteristic energy of the system, i.e. the Fermi energy, or, correspondingly, the Fermi temperature. Following from the arguments above, the paramagnetic configuration exhibits a smaller Fermi temperature when compared to the ferromagnetic configuration. This implies that the corresponding momentum distributions are smoothened more quickly which in turn lead to a relative increase of the kinetic energy of the paramagnetic configuration. This effect, hower, as we can see from Figure \ref{fig.applications.mbpt-fd-hf}, is not big enough to let the kinetic energy curves cross and they converge for $T\rightarrow\infty$.

\begin{figure}[t!]
  \includegraphics[width=\figurewidth]{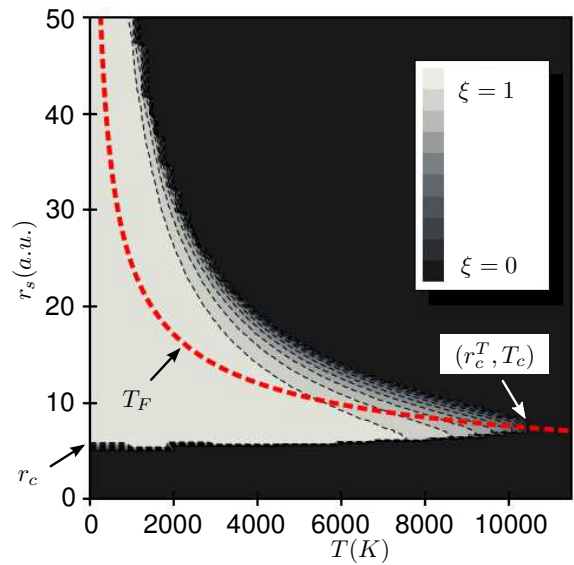}
  \caption{Phase diagram of the HEG for collinear spin configuration for the first-order FT-RDMFT functional. For $T<T_c$, when increasing $r_s$, the HEG shows both instantaneous and continuous phase transitions. The red dashed line denotes $T_F$.}
  \label{fig.heg.pd}
\end{figure}

The situation changes if one includes the exchange contribution and minimizes w.r.t. the momentum distribution. The exchange contribution for itself is known to be minimal for a ferromagnetic configuration with a momentum distribution describing a sharp Fermi sphere. Because it does not couple different spin channels this also implies that for general polarizations, the favourable configurations are the ones describing sharp Fermi spheres of appropriate radii. $\Omega_x$ therefore counteracts the effect of temperature which can be interpreted as an increase of the effective Fermi temperature $T_F^*$. As argued, this increase of $T_F^*$ is stronger for the ferromagnetic configuration, which, in addition to our considerations of noninteracting systems above, then leads to a cross-over of the kinetic energy curves in the \abref{FT-RDMFT} treatment. These arguments can be applied to both entropic as well as exchange contributions as well, but there the ferromagnetic configuration exhibits a lower value for zero temperature, preventing a crossing of energy curves. The behaviour of the free energy is more complicated because the entropy enters negatively. It has to be pointed out, however, that in the case of ``FD''-momentum distributions, i.e. in first order FT-MBPT, an increase of temperature first leads to an increase in the free energy before an eventually monotonic decrease. This is rather unphysical and can be appointed to the fact that the ``FD''-momentum distribution is not acquired by any sort of variational principle. The ``HF''-momentum distribution, one the other hand, is explicitly determined by a minimization procedure which leads to the qualitatively correct monotonical decrease of the free energy with temperature.

After these considerations, we now calculate the equilibrium polarization of the HEG for a wide range of densities and temperatures. The resulting magnetic phase diagram is shown in Figure \ref{fig.heg.pd}. The critical Wigner-Seitz radius $r_c$, marking the zero-temperature magnetic  transition between the paramagnetic and ferromagnetic phases has the well known value $5.56 a.u.$. On increasing the temperature, the ferromagnetic phase gets reduced until after some critical point it vanishes. The corresponding temperature $T_c$ is calculated to be at about $10500 K$ while the critical Wigner-Seitz radius $r_c^T$ becomes $7.1 a.u.$. Monte-Carlo results, on the other hand, show \cite{Ceperley_Alder.1980} that $r_c$ is expected to be about $75a.u.$. An investigation of the temperature dependence with the help of a Stoner model \cite{Zong_Lin_Ceperley.2002} gives $T_c\approx80K$ which is close to the corresponding Fermi temperature $T_F(r_s=75a.u.)=103K$. The fact that the critical temperature in the first-order, i.e. exchange only, treatment of FT-RDMFT, when compared to the Fermi temperature, is qualitatively correct suggests that the noninteracting entropy functional describes big parts of the interacting entropy correctly. The main focus in the development of more advanced functionals for FT-RDMFT should therefore be on the reproduction of the zero-temperature critical Wigner-Seitz radius. We therefore expect that if one uses a temperature-independent correlation functional which reproduces the zero-temperature properties of the HEG accurately, then the inclusion of the noninteracting entropy will yield a qualitatively correct phase diagram.

\begin{figure*}[t!]
  \centering
  \includegraphics[width=0.8\textwidth]{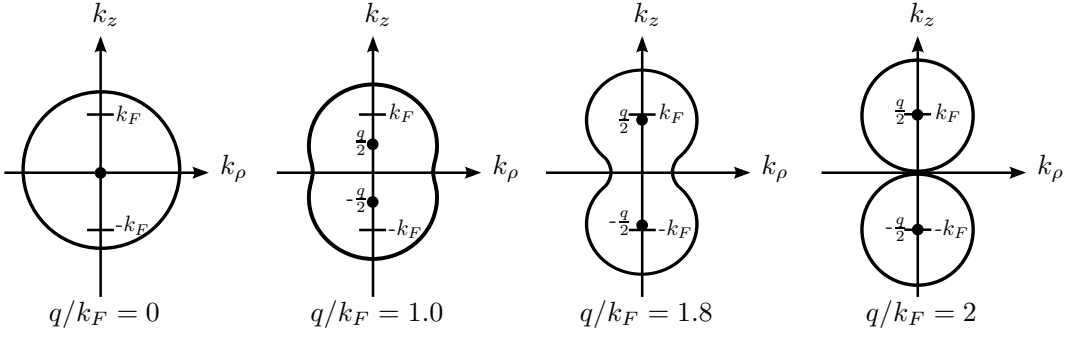}
  \caption{Sketch of $\tb q$-dependence of Fermi surface of a fully polarized \abref{PSS} state.}
  \label{fig.sdw.nk-q}
\end{figure*}

\subsection{Planar Spin Spirals}
Although collinear plane waves as NOs respect the symmetry of the system, they do not neccessarily yield the lowest free energy.
It was shown by Overhauser in 1962 \cite{Overhauser.1962} that for the electron gas at zero temperature in Hartree-Fock approximation, a state describing spin-density waves (SDW) or charge-density waves can yield energies below the symmetry preserving state. We will investigate how this changes for pure SDWs with increasing temperature of the electron gas. 

An investigation of SDWs within zero temperature RDMFT was done by Eich et al.\cite{Eich_Kurth_Proetto_Sharma_Gross.2010} and we will repeat the most important concepts in the following. The NOs describing a SDW also yields a spin channel seperable 1RDM and are given by
\begin{align}
  \phi_{\tb{k}1}(\tb{r})&=\left(
    \begin{array}{c}
      \cos\left(\frac{\Theta_k}{2}\right)e^{-i\tb{q}\cdot\tb{r}/2}\\
      \sin\left(\frac{\Theta_k}{2}\right)e^{i\tb{q}\cdot\tb{r}/2}
    \end{array}
  \right)\frac{e^{i\tb{k}\cdot\tb{r}}}{\sqrt{\Omega}}\\
  \phi_{\tb{k}2}(\tb{r})&=\left(
    \begin{array}{c}
      -\sin\left(\frac{\Theta_k}{2}\right)e^{-i\tb{q}\cdot\tb{r}/2}\\
      \cos\left(\frac{\Theta_k}{2}\right)e^{i\tb{q}\cdot\tb{r}/2}
    \end{array}
  \right)\frac{e^{i\tb{k}\cdot\tb{r}}}{\sqrt{\Omega}}.
\end{align}
The spin channels now do not exhibit spin-up or spin-down characteristics but describe spin spirals. To distinguish the SDW index from the collinear spin index, we will denote it by $b$ in the following. The free energy functional can now be written as
\begin{align}
  &F[\{n_{ib},\Theta_i\}]=\nn\\
  &\sum_{i,b}n_{ib}t_i+\frac{q^2}{8}-
  q\sum_i(n_{i1}-n_{i2})\cos(\Theta_i)Q_i\nn\\
  &-\frac12\sum_{i,j,b}(n_{ib}n_{bj})\cos^2\left(\frac{\Theta_i-\Theta_j}{2}\right)K_{ij}\nn\\
  &-\sum_{i,j,b}(n_{ib}n_{jb})\sin^2\left(\frac{\Theta_i-\Theta_j}{2}\right)K_{ij}\nn\\
  &+1/\beta\sum_{ib}\Big(n_{ib}\ln(n_{ib})+(1-n_{ib})\ln(1-n_{ib})\Big)\omega_i\label{eq.gp.sdw},
\end{align}
where
\begin{align}
  Q_i&=\frac{1}{2\rho}\int_{V_i}\frac{d^3k}{(2\pi)^3}k_z.
\end{align}
$t_i,K_{ij}$ and $w_i$ are given in Eqs. \eq{eq.weight.k}-\eq{eq.weight.w}.
The magnetization of the HEG for SDW NOs is given by
\begin{align}
  \tb{m}(\tb{r})&=-\left(
    \begin{array}{c}
      A\cos(\tb{q}\cdot\tb{r})\\
      A\sin(\tb{q}\cdot\tb{r})\\
      B
    \end{array}
  \right),
\end{align}
with the two amplitudes $A$ and $B$
\begin{align}
  A&=\frac12\int\frac{d^3k}{(2\pi)^3}(n_{\tb{k}1}-n_{\tb{k}2})\sin(\Theta_\tb{k})\label{eq.sdw.a}\\
  B&=\frac12\int\frac{d^3k}{(2\pi)^3}(n_{\tb{k}1}-n_{\tb{k}2})\cos(\Theta_\tb{k}).
\end{align}
In this work, we will restrict ourselves to planar SDWs, i.e. SDWs for which the magnetization of the HEG in z-direction vanishes ($B=0$). As argued in \cite{Conduit_Green_Simons.2009} and \cite{Eich_Kurth_Proetto_Sharma_Gross.2010} this planar configuration has a lower energy than a conical SDW with nonvanishing z-component. Furthermore, we will only consider SDWs with $\tb q\parallel \tb e_z$. These SDWs will be called planar spin spirals (PSS) in the following.

The requirement of vanishing z-component can be met by simple constraints on the ONs and NOs.
\begin{align}
  n_{(k_\rho,-k_z)b}&=n_{(k_\rho,k_z)b}\label{eq.pss.symm-1}\\
  \Theta_{(k_\rho,\pm|k_z|)}&=\frac{\pi}{2}(1\mp a_{(k_\rho,|k_z|)})\label{eq.pss.symm-2}
\end{align}
We can now minimize the free energy functional w.r.t. the ONs and orbital angles. To determine if at finite temperature a PSS state exhibits a lower free energy than the collinear states, we have to access the numerical values of these configurations in our PSS calculations. Using $N_1=\sum_in_{i1}$ and $N_2=\sum_in_{i2}$, we can define a PSS-polarization $\xi^{PSS}$ as
\begin{align}
  \xi^{PSS}&=\frac{N_1-N_2}{N_1+N_2}.
\end{align}
By setting $\tb q=0$ and $a_{k_\rho|k_z|}=1$, the PSS-NOs become
\begin{align}
  \phi_{(k_\rho,\pm|k_z|)1}(\tb{r})&=\frac{1}{\sqrt{2}}\left(
    \begin{array}{c}
      1\pm1\\
      1\mp1
    \end{array}
  \right)\frac{e^{i\tb{k}\cdot\tb{r}}}{\sqrt{\Omega}}\\
  \phi_{(k_\rho,\pm|k_z|)2}(\tb{r})&=\frac{1}{\sqrt{2}}\left(
    \begin{array}{c}
      -(1\mp1)\\
      1\pm1
    \end{array}
  \right)\frac{e^{i\tb{k}\cdot\tb{r}}}{\sqrt{\Omega}}
\end{align}
For this set of NOs, the PSS-unpolarized state $\xi^{PSS}=0$ corresponds to the paramagnetic state and the PSS-polarized one $\xi^{PSS}=1$ describes the ferromagnetic solution.

To understand the effect of an increase of $q$ it is instructive to consider a fully polarized noninteracting system. We show a sketch of the $q$-dependence of the Fermi surface in Figure \ref{fig.sdw.nk-q}. For $q=0$, the \abref{ON}s describe a Fermi sphere of radius $2^\frac13k_F$ around $k=0$. If one increases $q$, one can derive from the first three terms in Eq. \eq{eq.gp.sdw} and the symmetry relations \eq{eq.pss.symm-1}, \eq{eq.pss.symm-2} that the Fermi sphere will divide symmetrically along the $z$-direction. If $q$ supercedes $2k_F$, then there will be two distinct Fermi spheres with radius $k_F$, centered at $k=-q/2$ and $k=q/2$ respectively. 

If one now includes temperature effects, then the momentum distribution around the Fermi surface will be washed-out. To reproduce the paramagnetic collinear configuration one therefore would have to consider the limit $q\rightarrow\infty$. It is therefore possible to compare the free energies of a ferromagnetic collinear configuration ($q=0, \xi^{PSS}=1$), a fully polarized PSS configuration of finite $q$ and a paramagnetic collinear configuration ($q\rightarrow\infty, \xi^{PSS}=1$) by varying $q$ alone.

\begin{figure}[b!]
  \centering
  \includegraphics[width=\figurewidth]{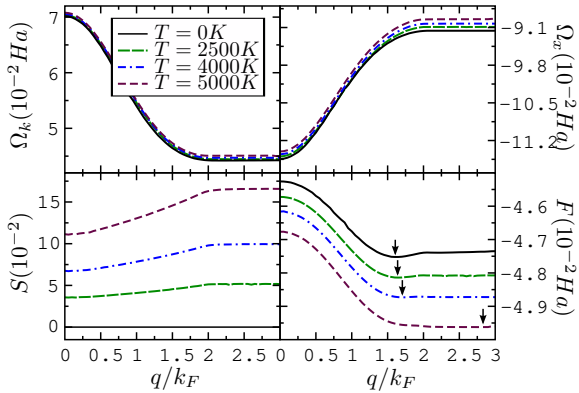}
  \caption{Clockwise from bottom left: entropy, kinetic energy, exchange energy, and free energy at $r_s=5.00a.u.$ $(T_F=23300K)$ versus the \abref{PSS} $\tb q$-vector. The minimal free energies are denoted by the arrows.}
  \label{fig.sdw.combined}
\end{figure}

\begin{figure}[b!]
  \centering
  \includegraphics[width=\figurewidth]{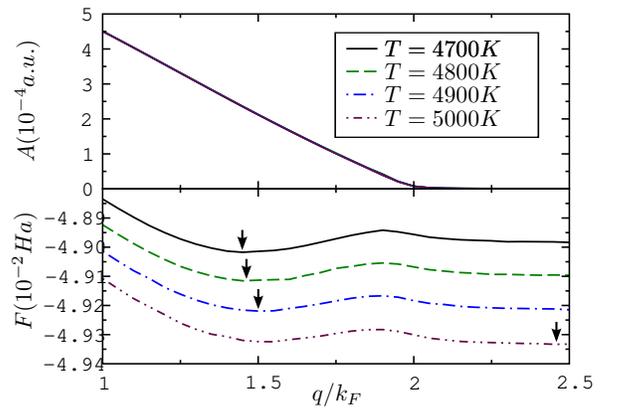}
  \caption{Amplitude $A$ and free energy $F$ for different $q$ and $T$ for $r_s=5.5a.u.$. The reduced temperatures are $t={0.244,0.249,0.254,0.260}$. The amplitudes change only slightly when increasing the temperature, but the optimal $q$ increases (arrows) discontinuously, letting the optimal amplitude $A_{opt}$ vanish. This denotes an instantaneous phase transition in the PSS phase.}
  \label{fig.sdw-first}
\end{figure}

The dependence of the entropy and the other thermodynamic variables on $q$ at several temperatures is shown in Figure \ref{fig.sdw.combined}. The entropy displays a monotonically increasing, almost linear dependence until the two Fermi spheres are completely seperated (at $q\approx2k_F$). This increase of the entropy in turn leads to an increase of the optimal $q$-vector (the $q$ for which the free energy is minimal).
This situation is depicted for $r_s=5.50$ in Figure \ref{fig.sdw-first} where we show the free energy and the amplitude of a PSS-state w.r.t. $q$ for several temperatures.
For the temperature below some critical temperature $T_c^{PSS}$, the free energy exhibits a minimum for finite $q$. If the temperature is increased above $T_c^{PSS}$, $q_{min}$ instantaneously jumps to a value bigger than $2k_F$, letting the amplitude of the PSS vanish. 
The jump in $q$ at $T_c^{PSS}$ therefore marks an instantaneous phase transition of the PSS phase where the amplitude of the optimal PSS is approaching zero instantaneously. In his work on spin density waves (SDW) in an electron gas, Overhauser \cite{Overhauser.1962} made the conjecture that on increasing the temperature the optimal amplitude of the SDW-state will approach zero continuously, giving rise to a continuous phase transition. Our results disprove the validity of this conjecture for PSS-states.

\begin{figure}[t!]
  \centering
  \includegraphics[width=\figurewidth]{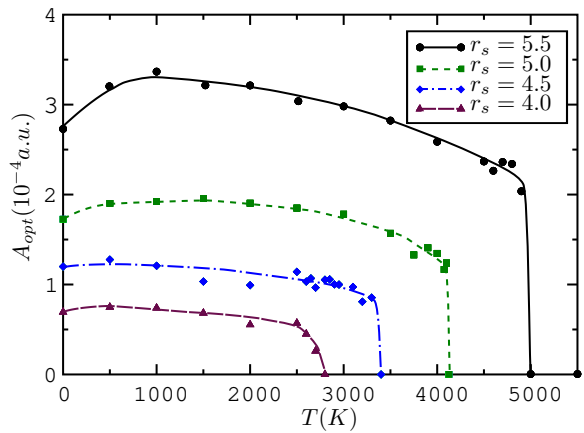}
  \caption{Amplitudes $A_{opt}$ of the quilibrium \abref{PSS} phases for different $r_s (\{4.0,4.5,5.0,5.5\}a.u.)$ and $T$. The corresponding Fermi temperatures are $\{36400,28800,23300,19200\}K$. The solid lines are mere guides to the eye.}
  \label{fig.sdw.amp-min}
\end{figure}

We show the dependence of the amplitude of the optimal PSS w.r.t $T$ for several $r_s$ in Figure \ref{fig.sdw.amp-min}.
For small $q$ the amplitude increases slightly when increasing the temperature which seems to be surprising at first. Temperature is usually expected to favour states of higher disorder and one could therefore expect the amplitude to be reduced.
However, in our calculations we encounter the following behaviour of the Fermi surface. For $q/k_F<2$ the zero-temperature Fermi surface assumes an hourglass-like shape as depicted in Figure \ref{fig.sdw.nk-q}. We find that an increase of the temperature smoothens the Fermi surface which in turn leads to an increase of occupation around the ``waist'' of the hourglass. The angles $\Theta_{\tb k}$ are usually closer to $\pi/2$ for smaller values of $k_z$ which then leads to an increase of the \abref{PSS} amplitude $A$ via Eq. \ref{eq.sdw.a}. We have to point out, however, the numerical uncertainty of these findings. The energy differences for slightly changed momentum distributions are very small but the amplitude shows a much stronger dependence. This also explains the big variance of the calculated optimal amplitudes, as shown in Figure \ref{fig.sdw.amp-min}. However, we did a careful minimization of the functional with several k-point-mesh refinements and believe that our numerical findings are qualtitatively correct.
We have argued that for small temperatures the amplitude of a \abref{PSS} state increases. The optimal $q$-vector, on the other hand, changes only slightly. Therefore, for small temperatures the optimal amplitude increases. It was already mentioned in \cite{Eich_Kurth_Proetto_Sharma_Gross.2010} that at zero temperature, the amplitude of the \abref{PSS} decreases with decreasing $r_s$. Below a critical radius $r_c^{PSS}=3.5$, we cannot resolve the amplitude of the \abref{PSS} anymore. This reduction of the amplitude is mainly due to the fact that the energy difference between the paramagnetic and ferromagnetic phases increases and therefore the optimal $\tb q$ approaches $2k_F$ (see Figure \ref{fig.sdw-first}). When considering finite temperatures, this leads to a decrease of the critical temperature $T_c^{PSS}$, as shown in Table \ref{table.sdw.amp}.

\begin{table}[b!]
  \centering
    \begin{tabular}{|x{17mm}|x{10mm}|x{10mm}|x{10mm}|x{10mm}|}
    \hline
    $r_c(a.u.)$&4.0&4.5&5.0&5.5\\
    \hline
    $T_c(K)$&2700&3400&4100&5000\\
    \hline
  \end{tabular}
  \caption{Critical temperatures $T_c^{PSS}$ above which no equilibrium PSS-phase was found.}\label{table.sdw.amp}
\end{table}
For those $r_s$, for which a polarized configuration is favourable over an unpolarized one, a formation of a PSS was found to increase the free energy.

In Figure \ref{fig.pd.combo} we combine the results from the collinear-NOs and the PSS-NOs calculations to get a schematic diagram of the magnetic phase diagram of the HEG.
\begin{figure}[t!]
  \caption{Phase diagram of the HEG in first-order-/HF- FT-RDMFT approximation. FM: ferromagnetic collinear phase, PM: paramagnetic collinear phase, PP: partially polarized  collinear phase, PSS: planar spin spiral state. The thick dashed red line denotes instantaneous phase transitions. The dotted blue line marks the Fermi temperature $T_F$.}
  \label{fig.pd.combo}
\end{figure}

Considering the collinear phases, the kinetic energy favours a paramagnetic configuration whereas the exchange contribution is minimized for a ferromagnetic one. At zero temperature, the different $r_s$-dependencies lead to a magnetic phase transition at $r_c$. Inclusion of the entropy, which itself favours a paramagnetic configuration, leads to a weakening and eventually vanishing ferromagnetic phase with increasing temperature. This behaviour and the decrease of the kinetic energy w.r.t. the polarization for high temperatures could also be explained by the concept of ``{}effective''{} Fermi temperatures. The magnetic phase transition between paramagnetic and polarized phases for increasing $r_s$ was shown to be instantaneous while the transition from polarized to paramagnetic phases for higer $r_s$ is second-order. As can be seen from Figure \ref{fig.pd.combo}, the continuous phase transitions occur close to the Fermi temperature $T_F$. The PSS phase shows a much stronger density dependence and is strongest for $4\leq r_s\leq6$. In this range we also see a temperature-driven instantaneous phase transition between a PSS state and the paramagnetic state. The phase diagram from first-order FT-RDMFT agrees qualitatively with the one derived in Ref. \cite{Conduit_Green_Simons.2009}  employing a field-theoretical approach in combination with a contact interaction approximation. This method also yields a favourable PSS phase in the paramagnetic to ferromagnetic phase transition at a fixed temperature, given that this temperature is below some critical value. This suggests that correlation functionals in FT-RDMFT should preserve this general property.

This concludes our investigation of the phase diagram of the \abref{HEG} for the first-order \abref{FT-RDMFT} functional.

We have shown before \cite{Baldsiefen_al_1.2012} that the minimization of this first-order functional is equivalent to a solution of the \abref{FT-HF} equations. In the following, we will therefore investigate the temperature dependence of the \abref{FT-HF} single-particle dispersion relation for collinear as well as for \abref{PSS} configurations.
\begin{figure}[t!]
\centering
  \includegraphics[width=\figurewidth]{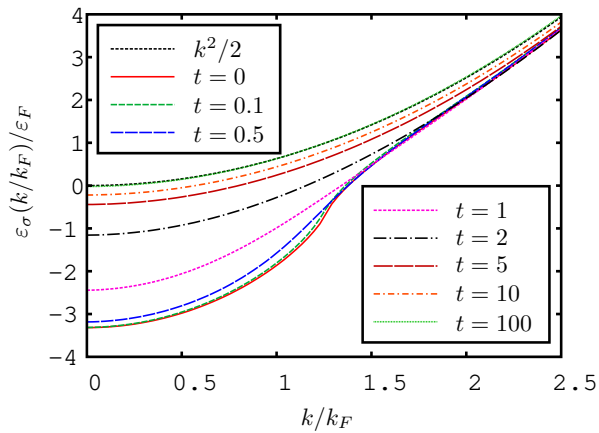}
  \caption{Hartree-Fock dispersion relation for the \abref{HEG} in collinear spin configuration at $r_s=5a.u.$ for different values of the effective temperature $t=T/T_F$. The black dotted line denotes the noninteracting relation $\veps(k)=k^2/2$.}
  \label{fig.hf_disp}
\end{figure}
\subsection{Dispersion relations}\label{sec.applications.hf-disp} 
As is well known, the \abref{HF} dispersion relations for collinear spin configuration at zero temperature can be calculated analytically to yield
\begin{align}
  \veps_\sigma(k)&=\frac{k^2}{2}-\frac{2}{\pi}f_d(k/k_{F\sigma}).
\end{align}
$f_d(x)$ denotes the dimension-dependent corrections. In three dimensions it reads
\begin{align}
  f_3(x)&=\frac12+\frac{1-x^2}{4x}\log\left|\frac{1+x}{1-x}\right|.
\end{align}

At finite temperature, because of the entropy, the minimum of the free energy functional will not exhibit pinned states. Therefore, at the minimum of $F[\{n_i\}]$ from Eq. \eq{eq.heg.f}, the following relation holds
\begin{align}
  \frac{\partial F}{\partial n_i}&=0=\frac{\partial \Omega_k}{\partial n_i}+\frac{\partial \Omega_x}{\partial n_i}-\frac1\beta\frac{\partial S_0}{\partial n_i}.
\end{align}
With the explict form of $S_0[\{n_i\}]$ of Eq. \eq{eq.heg.s0} this yields
\begin{align}
  \veps_i-\mu&=\frac{\partial \Omega_k}{\partial n_i}+\frac{\partial \Omega_x}{\partial n_i}\label{eq.applications.ftrdmft.e_hf},
\end{align}
which can easily be accessed numerically. FT-RDMFT can therefore be used to efficiently investigate temperature effects on the HF-eigenenergies, e.g. the closing of the energy gap between spin-channels in either collinear or chiral spin configurations. We have chosen a density at which the zero-temperature groundstate is given by a PSS state ($r_s=5a.u.$) and calculated the \abref{FT-HF} dispersion relation for the fully polarized HEG for various values of $t$. The results are shown in Figure \ref{fig.hf_disp} where, because of the full polarization, the Fermi surface is at $k_{F\uparrow}=2^{\frac13}k_F$. As the exchange contribution is quadratic in the ONs and does not couple the different spin channels in collinear configuration (see Eqs. \eq{eq.heg.coll.x},\eq{eq.heg.ex}), the unoccupied band shows the noninteracting dispersion relation $k^2/2$. For the occupied band at zero temperature, we recover the exact relations to good agreement. An increase of the temperature then firstly lets the ``valley'' in the dispersion relation vanish and then secondly lets the eigenenergies approach the noninteracting relation, therefore closing the \abref{HF}-gap. This can be understood from the thermal smoothening of the momentum distribution, induced by the entropy. As \abref{NO}s of very different quantum numbers generally have only small overlap, this leads to a decrease of the influence of the exchange contribution compared to the kinetic energy.

\begin{figure}[t!]
\centering
  \includegraphics[width=\figurewidth]{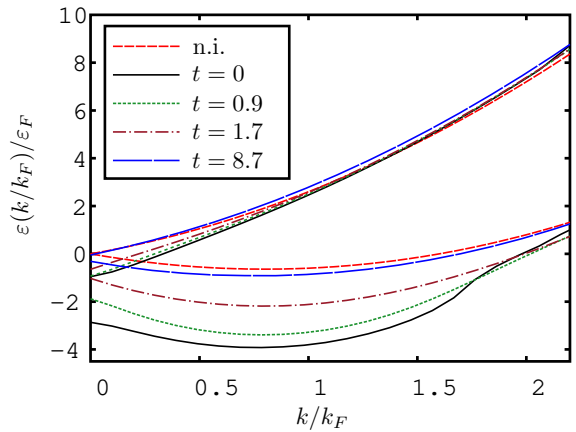}
  \caption{Hartree-Fock dispersion relation for the PSS phase in the \abref{HEG} at $r_s=5.00a.u.$ $(T_F=23300K)$ and $q=1.6k_F$. The red medium-dahed lines (n.i.) denote the noninteracting dispersions $e_b^0(k)$ whereas the other lines show the dispersion relation for a ferromagnetic configuration for several reduced temperatures $t$. }
  \label{fig.sdw_disp}
\end{figure}

As a second choice for the \abref{NO}s of the \abref{1RDM} we will now consider the PSS states. The single-particle eigenenergies can be determined in the same way as for the collinear case (see Eq. \eq{eq.applications.ftrdmft.e_hf}). The groundstate at $r_s=5a.u.$ is given by a fully polarized \abref{PSS} configuration with $q=1.6k_F$, which leads to the dispersion relation shown in Figure \ref{fig.sdw_disp}. The noninteracting \abref{PSS} energies $e_b^0(k)$ are simply given by
\begin{align}
  e_b^0(k)&=\frac12(k^2\pm kq)=\frac{\left(k\pm\frac{q}{2}\right)^2}{2}-\frac{q^2}{8}.
\end{align}
As a first difference to the collinear case, we see that the unoccupied band at zero temperature does not follow the noninteracting relation. This is due to the fact that the exchange contribution in Eq. \eq{eq.gp.sdw} explicitly couples both \abref{PSS} channels.  As for the collinear configuration, the temperature first lets the ``valley'' in the dispersion relation of the occupied band disappear. With a further increase of the temperature, both unoccupied as well as occupied bands then approach the noninteracting dispersion relation. The fact that for both collinear as well as PSS configurations the kink in the dispersion relation disappears much faster than the overall convergence to the noninteraction relation can serve as an argument, why for intermediate temperatures an effective mass approximation might be feasible.

\section{Conclusions}
In Part I of this work \cite{Baldsiefen_al_1.2012} we have introduced a new theoretical framework for the description of thermodynamic variables of quantum systems in grand canonical equilibrium using the 1RDM of the system. By employing methods from FT-MBPT this framework allows an, in principle exact, construction of the grand potential functional.
As a first utilization of this theory we apply the corresponding first-order functional, whose minimization we have shown \cite{Baldsiefen_al_1.2012} to be equivalent to the solution of the finite-temperature Hartree-Fock equations, to the HEG. We consider both the collinear spin phase and the PSS phase, possessing a chiral spin symmetry.

When compared to a first-order treatment in FT-MBPT, FT-RDMFT removes the unphysical increase of the free energy with increasing temperature. This can be understood from the fact that FT-MBPT yields an approximation to the Green's function. This only ensures that approximate results are close to the real ones but it does not ensure the correct behaviour with respect to external parameters such as the temperature.
FT-RDMFT, on the other hand, determines the 1RDM via a minimization procedure, which in the case of the first order functional ensures a decrease of the free energy with temperature.

The collinear phase diagram exhibits both instantaneous as well as continuous phase transitions. When calculating the amplitude of the optimal PSSs we see the expected decrease with increasing temperature. However, in contrast to Overhauser's proposition, this decrease contains a nonanlytical jump in the amplitude of the PSS, marking an instantaneous phase transition. Subsequently, we have employed FT-RDMFT to investigate the temperature dependence of the Hartree-Fock dispersion relations for both collinear as well as PSS phases. We have found that the influence of temperature reduced the characteristic kink in the dispersion relation rather quickly whereas the assumption of the limiting noninteraction dispersion proceeds on a much longer temperature scale.

Following from our treatment of the first-order functional of FT-RDMFT, the next step will the derivation of approximate functionals for the correlation part of the grand potential. Part III \cite{Baldsiefen_al_3.2012} of this work focusses on this task and will present several ways to approach this problem.


\end{document}